\documentclass[5p]{elsarticle}

\usepackage{natbib}
\usepackage{graphics}
\usepackage{graphicx}
\usepackage{epsfig}
\usepackage{color}
\usepackage{slashed}
\usepackage{multirow}
\usepackage{tabularx,ragged2e,booktabs,caption,fullpage}

\usepackage{bm}
\usepackage{footnote}

\usepackage{amssymb, amsmath, latexsym, fullpage}
\usepackage{eqnarray}

\allowdisplaybreaks

\newcommand{\la}{\langle}
\newcommand{\ra}{\rangle}

\newcommand{\msun}{M_{\odot}}

\usepackage[bottom,flushmargin,hang,multiple,symbol]{footmisc}

\usepackage[colorlinks=true,linkcolor=blue,citecolor=blue]{hyperref}

\begin{document}

\begin{frontmatter}
















\title{Strangeness in Neutron Star Cooling}

\author[label1]{Yeunhwan Lim}
\ead{ylim9057@ibs.re.kr}
\address[label1]{Rare Isotope Science Project,
Institute for Basic Science,
Daejeon 34047, Republic of Korea}

\author[label2]{Chang Ho Hyun}
\ead{hch@daegu.ac.kr}
\address[label2]{Department of Physics Education,
Daegu University,
 Gyeongsan 38453, Republic of Korea}


\author[label3]{Chang-Hwan Lee}
\ead{clee@pusan.ac.kr}
\address[label3]{Department of Physics,
Pusan National University, 
Busan 46241, Republic of Korea}

\begin{abstract}
We study the thermal evolution of neutron stars in the presence of hyperons or kaons in the core. 
Our results indicate that the nucleon and hyperon direct Urca processes play crucial roles for the cooling of neutron stars.
The presence of hyperons drives fast cooling mechanisms in two ways:
1) it allows the hyperon direct Urca prior to the nucleon direct Urca, 
2) and it makes the nucleon direct Urca more feasible by reducing the neutron Fermi momentum.
We found that the neutron star equation of state (EOS) with hyperons can be consistent with both mass and temperature observations. 
We also found that the neutron star EOS with kaon condensation can be consistent with observations, 
even though the cooling behavior is seldom useful to identify or isolate the effect of kaon condensation.
\end{abstract}





\end{frontmatter}

\def\chlee#1{\marginpar[$\Rightarrow$]{$\Leftarrow$}{\bf \em CHL: #1}}


\section{Introduction}\label{Sec:int}

One of the most interesting stellar objects for nuclear astrophysics in the Universe
is a neutron star. 
It is not only an astrophysical object to be observed,
but also a testing probe for high-density nuclear matter which cannot exist
on earth in a natural environment.
It is believed that the baryon number density at the center of neutron stars reaches  several times of 
normal nuclear matter saturation density ($n_0 = 0.16\,\rm{fm}^{-3}$). 
As the baryon number density increases, various new degrees of freedoms, such as hyperons, kaons, and even deconfined quarks, 
can come into the nuclear matter.
The inner structure of neutron star is still unknown
mainly because of the theoretical uncertainties in the extrapolation of dense matter physics to high densities~\cite{lattimer2015}.

Observations on neutron stars are mainly focused on the macroscopic properties such as masses and radii.
By observing the macroscopic properties, it's not easy to investigate the inner structure of neutron stars 
because multiple EOSs can give similar results for the mass and radius.
Neutron star's temperature provides additional information that can help us understand the inner structure of neutron stars.
Since the dominant contribution to the cooling of a neutron star at a given 
time strongly depends on the inner structure, by combining multiple observations of 
neutron stars at various ages, one may be able to identify EOS 
which is consistent with neutron star cooling.

In this work, we try to understand the thermal evolution of neutron stars
with the nuclear EOSs which include hyperons or kaons at high densities.
We use non-relativistic nuclear models to construct nucleonic degrees of freedom and
employ non-relativistic hyperon potential models for $N \Lambda$ and $\Lambda \Lambda$ interactions.
We combine non-relativistic nuclear models and SU(3) non-linear chiral model for the description of kaon interactions. 
For the cooling calculation, in addition to the standard cooling process, 
we include direct Urca processes for both nucleons and hyperons  and the neutrino emission caused by the condensed kaons.
We neglect the nucleon superfluidity effects because its contribution is not crucial in explaining the observations 
once the nucleon direct Urca process sets in as discussed in Lim {\it et al.}~\cite{lim2015a}.
We also neglect the $\Lambda$ superfluidity effect because the composition of hyperons is highly model dependent and
the pairing between $\Lambda$ hyperons is not certain.

This paper is organized as follows. In Sec. \ref{Sec:str}, we briefly summarize
the nuclear models with hyperons and kaons.
In Sec. \ref{Sec:res}, we show the result of cooling simulation in which the nuclear EOSs
with strangeness are used. In Sec \ref{Sec:con}, we summarize
the results and discuss their implication to the EOS for high density nuclear matter.

\section{Strangeness in nuclear matter}\label{Sec:str}

\subsection{$\Lambda$ hyperons and neutrino emission}

In order to describe the dense matter with hyperons, 
the Skyrme nucleon-nucleon interactions SLy4 \cite{sly4}, SkI4 \cite{ski4}, and
SGI \cite{sgi} are combined with non-relativistic hyperon potential models.
In Table~\ref{tb:nsmass}, we summarize the maximum of neutron stars with these models and the critical mass for the nucleon direct Urca process.
Note that, with only nucleons, SLy4 model does not allow the direct Urca process because the proton fraction is not sufficient enough~\cite{lim2015a}.

\begin{table}[t]
\caption{Maximum mass of neutron stars in units of the mass of Sun ($M_\odot$) 
for each nuclear model without strangeness. The number in
parenthesis indicates the critical mass of the neutron star for nucleon direct Urca
process (involving electrons). }
\begin{center}
\begin{tabular}{cccc}
\\[-6.0ex]
\toprule
         &   SLy4              &  SkI4              &  SGI \\
\hline
\\[-2.0ex]
$ M_{\text{max}} $    & 2.07 (-)          & 2.19 (1.63)        & 2.25 (1.72) \\
\bottomrule
\end{tabular}\label{tb:nsmass}
\end{center}
\end{table}

In Ref.~\cite{lim2015}, we showed that proper combinations of Skyrme forces for the $NN$,
$N\Lambda$, and $\Lambda\Lambda$ interaction can produce the maximum mass of
a neutron star close to $2\msun$ even if hyperons soften the EOS.
In the non-relativistic Skyrme force model, 
the potentials for the $N\Lambda$ and $\Lambda\Lambda$ interactions take the form
\begin{equation}
v_\Lambda = v_{N \Lambda} \delta(\bm{r}_{N\Lambda}) + v_{\Lambda\Lambda} \delta(\bm{r}_{\Lambda\Lambda}),
\end{equation}
where
\begin{align}
v_{N\Lambda} (\bm{r}_{N\Lambda})  =
      &\ u_0 (1+y_0 P_\sigma)   
        +\frac{1}{2} u_1 \left[ \overleftarrow{\bm{k}}^2_{N\Lambda} 
         +    \overrightarrow{\bm{k}}^2_{N\Lambda}\right]  \nonumber \\
      & + u_2 \overleftarrow{\bm{k}}_{N\Lambda} \cdot  \overrightarrow{\bm{k}}_{N\Lambda} \nonumber \\
      & + \frac{3}{8} u'_3 (1+y_3 P_\sigma) \rho^\gamma_N
             \left(\frac{\bm{r}_N+\bm{r}_\Lambda}{2}\right) 
              , \nonumber \\
v_{\Lambda\Lambda} (\bm{r}_{ij}) =  
       & \  \lambda_0   + \frac{1}{2} \lambda_1 \left[ \overleftarrow{\bm{k}}^2_{ij}  
            +  \overrightarrow{\bm{k}}^2_{ij}\right] \nonumber \\ 
       & + \lambda_2 \overleftarrow{\bm{k}}_{ij} \cdot   \overrightarrow{\bm{k}}_{ij} 
            + \lambda_3 \rho^\alpha_N (\bm{R}).
\end{align}
$P_\sigma$ is the spin exchange operator, 
$\delta(\bm{r}_{ij}) = \delta(\bm{r}_i-\bm{r}_j)$,
$\bm{R}=(\bm{r}_i+\bm{r}_j)/2$, 
$\bm{k}_{ij} =  \bm{k}_i -  \bm{k}_j$, and
the left (right) arrows indicate the momentum operators applied to the left (right).
The parameters of $N\Lambda$ interactions are employed from Ref.~\cite{gul2012} (HP$\Lambda$2) 
and Ref.~\cite{ybz1988} (YBZ6),
and those of $\Lambda\Lambda$ interactions from Ref.~\cite{lans1998} (S$\Lambda\Lambda$3) 
and Ref.~\cite{minato2011} (S$\Lambda\Lambda$3$^\prime$).
In Table~\ref{tb:nshyper}, the masses of neutron stars with $\Lambda$ hyperons are summarized.
The lower limit is the mass at which $\Lambda$ hyperons start to contribute and the upper limit is the
maximum mass of neutron stars in the presence of $\Lambda$ hyperons.

\begin{table}[t]
\caption{Mass of neutron stars with $\Lambda$ hyperons. 
Masses are given in units of $M_{\odot}$.
For the $N\Lambda$ interaction, HP$\Lambda2$ model is used for SLy4, and YBZ6 model for SkI4 and SGI.
For the $\Lambda\Lambda$ interaction, S$\Lambda\Lambda3$ model is combined with SLy4 and SGI,
and S$\Lambda\Lambda3'$ with SkI4. }
\begin{center}
\begin{tabular}{ccc}
\\[-6.0ex]
\toprule
  SLy4              &  SkI4              &  SGI \\
\hline
\\[-2.0ex]
 $1.15 \sim 1.85$  &  $1.47 \sim 1.97 $ & $1.44 \sim 2.04 $   \\
\bottomrule
\end{tabular}\label{tb:nshyper}
\end{center}
\end{table}

The hyperon direct Urca process we are interested in is the one involving $\Lambda$,
\begin{align}
\Lambda & \rightarrow p + l + \bar{\nu}_l \, , \nonumber \\
p + l & \rightarrow \Lambda + \nu_l\,,
\end{align}
where $l$ stands for leptons.
The emission rate of $\Lambda$ hyperon direct Urca process was obtained by Prakash {\it et al.} \cite{prakash1992},
\begin{equation}
\begin{aligned}
Q_\Lambda = & 4.0\times 10^{27} 
\frac{m_\Lambda^{*}m_p^*}{m_\Lambda m_p}
\left(\frac{n_e}{n_0}\right)^{1/3}\, R\  T_9^{6}\,\\
& \times \Theta_t\, 
  \mathrm{erg}\, \mathrm{cm}^{-3}\,\mathrm{s}^{-1}\,,
\end{aligned}
\end{equation} 
where $T_9 = T / 10^9 \mathrm{K}$, $\Theta_t$ is a step function taking into account the 
momentum conservation condition, and $R$ $(\approx 0.0394)$ is the relative ratio
of the hyperon direct Urca process to the nucleon direct Urca process. 
We note that even though $R$ is almost two orders of magnitude smaller than unity,
the $\Lambda$ hyperon direct Urca process is more powerful than the modified Urca process 
($Q_{\rm mUrca}\sim 10^{21}\,T_9^8\, \mathrm{erg}\, \mathrm{cm}^{-3}\,\mathrm{s}^{-1}$).
The $\Lambda$ hyperon direct Urca process happens right after the
appearance of $\Lambda$ in the core of neutron stars. For example, 
in the SLy4+HP$\Lambda2$+S$\Lambda\Lambda3$ model, the critical density
for hyperon appearance is $n = 0.453$ fm$^{-3}$ and the threshold density for the
$\Lambda$ hyperon direct Urca process is $n = 0.454$ fm$^{-3}$. 
Thus, we can regard that the hyperon direct Urca process is ready to occur
as soon as the $\Lambda$ hyperon exists.
Table \ref{tb:nsurca} shows the threshold mass of neutron stars both for
the nucleon and the hyperon direct Urca processes in each model. 
The threshold mass for the hyperon direct Urca process is lower than the one for the
nucleon direct Urca process in each case.

\begin{table}[t]
\caption{The critical mass of neutron stars for the hyperon (nucleon) direct Urca
process in the presence of hyperons.
Combination of the $N\Lambda$ and $\Lambda\Lambda$ interactions are the same
with Table~\ref{tb:nshyper}.
Numbers are given in units of $M_\odot$.
}
\begin{center}
\begin{tabular}{ccccc}
\\[-6.0ex]
\toprule
SLy4              & SkI4            &        SGI      \\
\hline
1.16 (1.85)       &  1.49 (1.63)     &     1.45 (1.64) \\
\bottomrule
\end{tabular}\label{tb:nsurca}
\end{center}
\end{table}

\subsection{Kaon condensation and neutrino emission}

In order to describe the dense matter with kaons, 
the Skyrme nucleon-nucleon interactions SLy4 \cite{sly4}, SkI4 \cite{ski4}, and
SGI \cite{sgi} are combined with an SU(3) non-linear chiral model suggested by Kaplan and Nelson~\cite{kaplan1986}. 
As discussed in Lim {\it et al.}~\cite {lim2014}, one can obtain the kaonic part of the interaction 
from the effective Lagrangian density given as
\begin{align}
{\cal L}_K = & f^2_\pi \frac{\mu_K^2}{2}\sin^2\theta_K 
- 2m^2_K f^2_\pi\sin^2\frac{\theta_K}{2} \nonumber \\ 
   & + (n^\dagger n + 2 p^\dagger p) \  \mu_K \sin^2 \frac{\theta_K}{2} \nonumber \\
   & - p^\dagger p \ 2 a_1 m_s \sin^2 \frac{\theta_K}{2} \nonumber \\
   & - (n^\dagger n + p^\dagger p)  (2 a_2 + 4 a_3) m_s  \sin^2 \frac{\theta_K}{2} ,
   \label{Klagrangian}
\end{align}
where $f_\pi$ ($\simeq 93$~MeV) is the pion decay constant, 
$\mu_K$ the kaon chemical potential,  
$a_1 m_s \simeq -67$~MeV and  $a_2 m_s \simeq 134$~MeV. 
The value of $a_3 m_s$ is related to the kaon-nucleon sigma term which is still uncertain. 
We consider three values of $a_3 m_s$, $-134$, $-178$, and $-222$ in MeV, 
which correspond to the strangeness contents of the proton $\langle \bar s s \rangle$ = 0, 0.01, and 0.1, respectively. 
In the mean field approximation, the baryon densities, $\langle p^\dagger p\rangle =\rho_p$ and 
$\langle n^\dagger n \rangle = \rho_n$ are coupled to the kaon field as in Eq.~(\ref{Klagrangian}), 
where the condensed kaons are parameterized by the expectation value of kaon field $\langle K^-\rangle$,
\begin{equation}
\theta_K \equiv \sqrt 2 \frac{|\langle K^- \rangle|}{f_\pi}.
\end{equation}

The maximum mass of neutron stars for each nuclear model without strangeness 
is summarized in Table~\ref{tb:nsmass}. 
Note that we choose the models which are consistent with the observations of 
$2\msun$ neutron stars \cite{demorest2010, Anto13}.
In Table~\ref{tb:nskaon}, 
the masses of neutron stars with kaon condensation are summarized for three values of $a_3 m_s$.
The lower limit is the mass at which the kaon condensation starts to appear in the core, and the upper limit 
is the maximum mass of neutron stars with kaon condensation. Note that, due to the condensed kaons, 
the maximum masses are reduced by $3\sim 15\%$ depending on the model, and $a_3 m_s = -222$~MeV is 
not consistent with the current observation of $2\msun$ neutron stars.

\begin{table}[t]
\caption{Mass of neutron stars with kaon condensation \cite{lim2014}.
Masses are in units of $M_{\odot}$ and the values of $a_3m_s$ are given in units of MeV.  }
\begin{center}
\begin{tabular}{cccc}
\\[-6.0ex]
\toprule
$a_3m_s$ &   SLy4              &  SkI4              &  SGI \\
\hline
\\[-2.0ex]
$-134$ & $1.94 \sim 1.99$  &  $2.00 \sim 2.06 $ & $2.12 \sim 2.19 $   \\
$-178$ & $1.74 \sim 1.83$  &  $1.84 \sim 1.90 $ & $1.95 \sim 2.03$  \\
$-222$ & $1.49 \sim 1.79 $ &  $1.64 \sim 1.85 $ &  $1.75 \sim 1.93 $   \\
\bottomrule
\end{tabular}\label{tb:nskaon}
\end{center}
\end{table}

Brown {\it et al.} \cite{brown1998} calculated the neutrino emission rate
with the condensed kaons,
\begin{equation}
n + \la K^{-} \ra \rightarrow n + l + \bar{\nu}_l \,.
\label{eq:kurca}
\end{equation}
The emission rate is given by
\begin{equation}
\begin{aligned}
Q_K = & 2.5\times 10^{26} \frac{m_n^{*2}}{m_n^2}
\left(\frac{n_e}{n_0}\right)^{1/3}\,T_9^{6}\,\theta_K^2\\
& \times \tan^2\theta_C\, 
  \mathrm{erg}\, \mathrm{cm}^{-3}\,\mathrm{s}^{-1}\, ,
\end{aligned}
\end{equation}
where $m_n^*$ is the neutron effective mass, $n_e$ is the electron number density,
and $\theta_C$ represents Cabbio angle. 
The emission rate is doubled if both electrons and muons exist~\cite{yakovlev00}.

\section{Neutron star cooling curves}
\label{Sec:res}

\subsection{$\Lambda$ hyperons}

\begin{figure}[t]
\begin{center}
\includegraphics[scale=0.4]{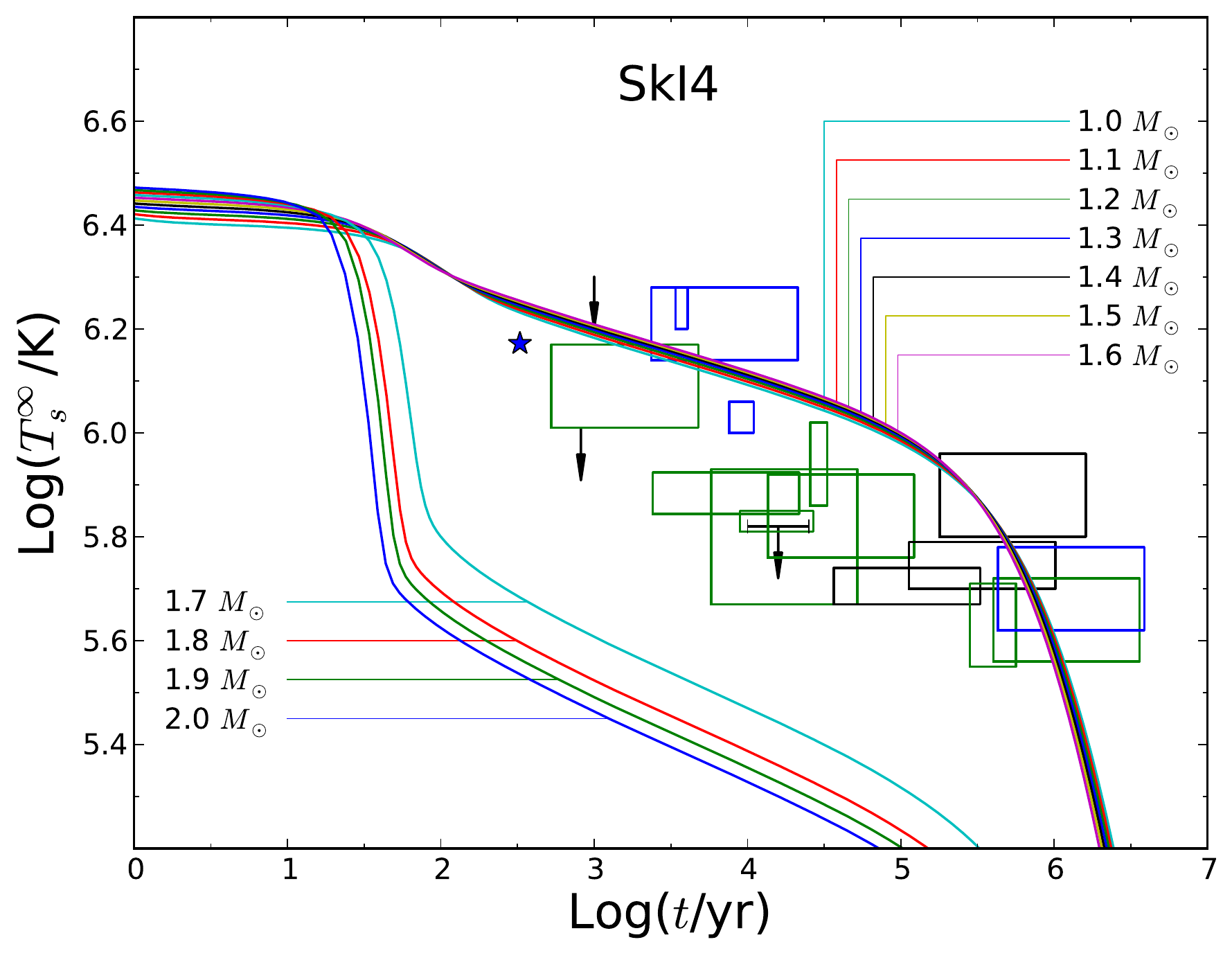}\\
(a) \\[10pt]
\includegraphics[scale=0.4]{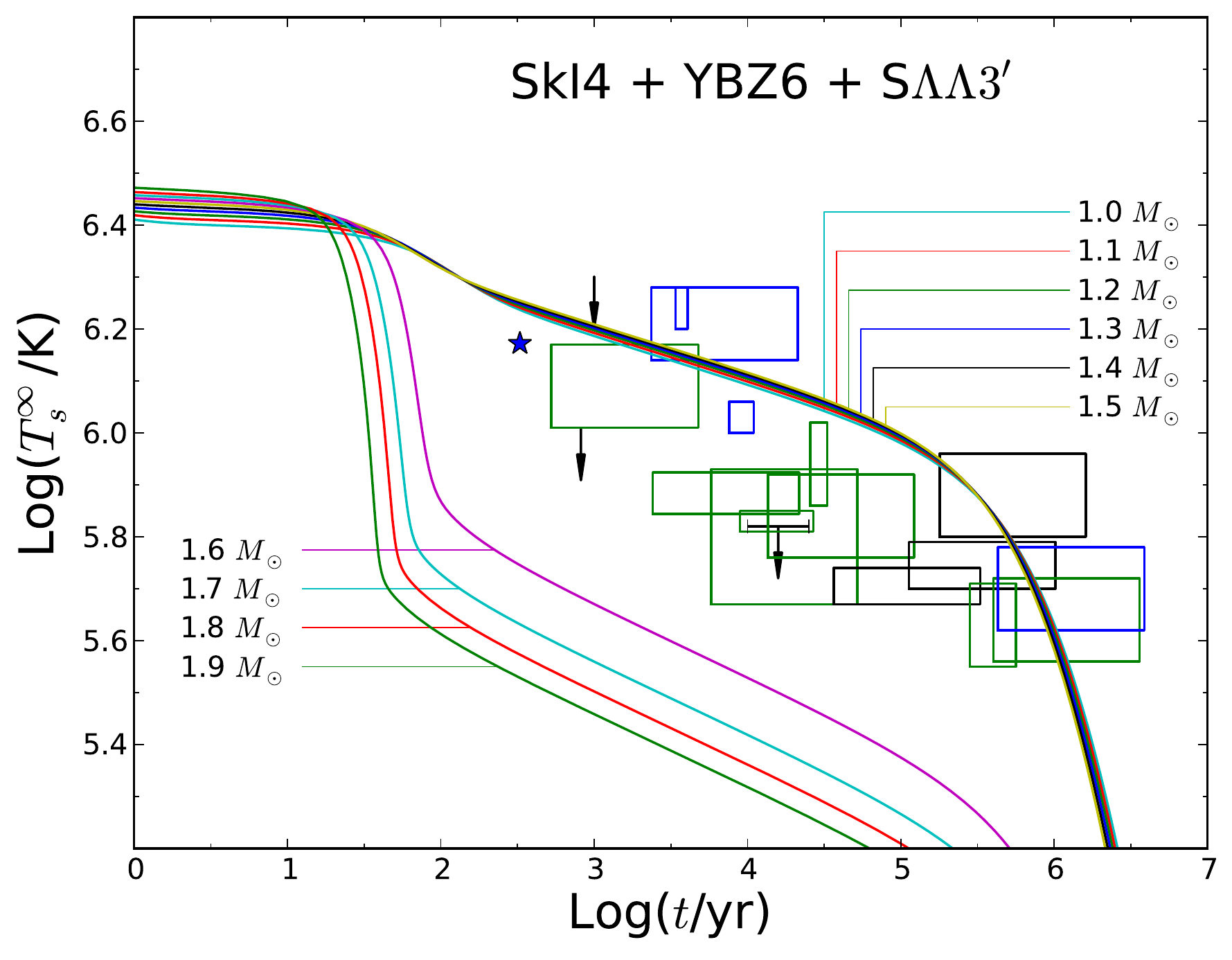}\\
(b)
\end{center}
\caption{Cooling curves (a) without strangeness~\cite{lim2015a} and (b) with hyperons for the combined model SkI4+YBZ$6$+S$\Lambda\Lambda 3^\prime$. 
The boxes indicate the ranges of estimated temperatures of 19 observed neutron stars.
}
\label{fig:ski4hyper}
\end{figure}

\begin{figure}[t]
\begin{center}
\includegraphics[scale=0.4]{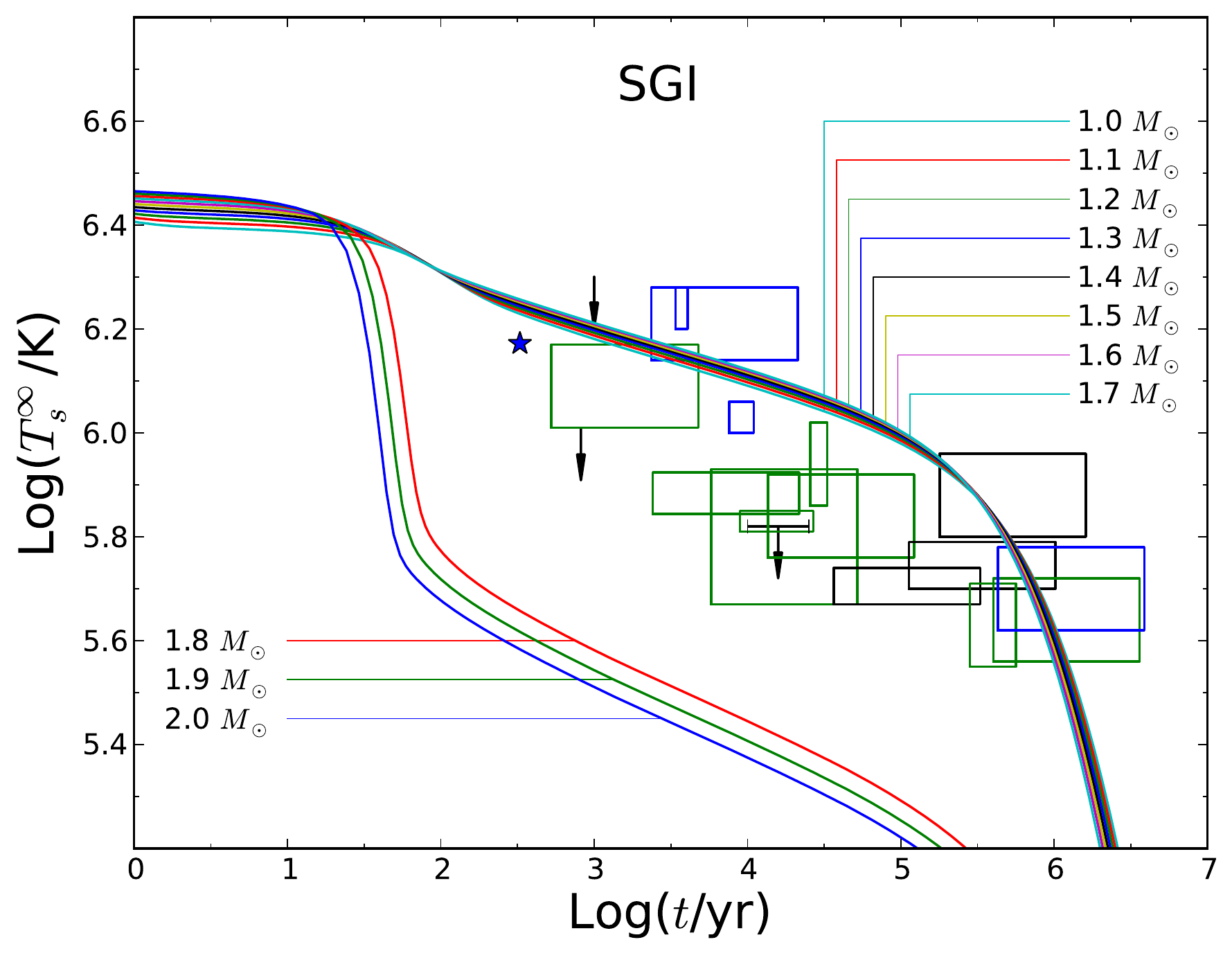}\\
(a) \\[10pt]
\includegraphics[scale=0.4]{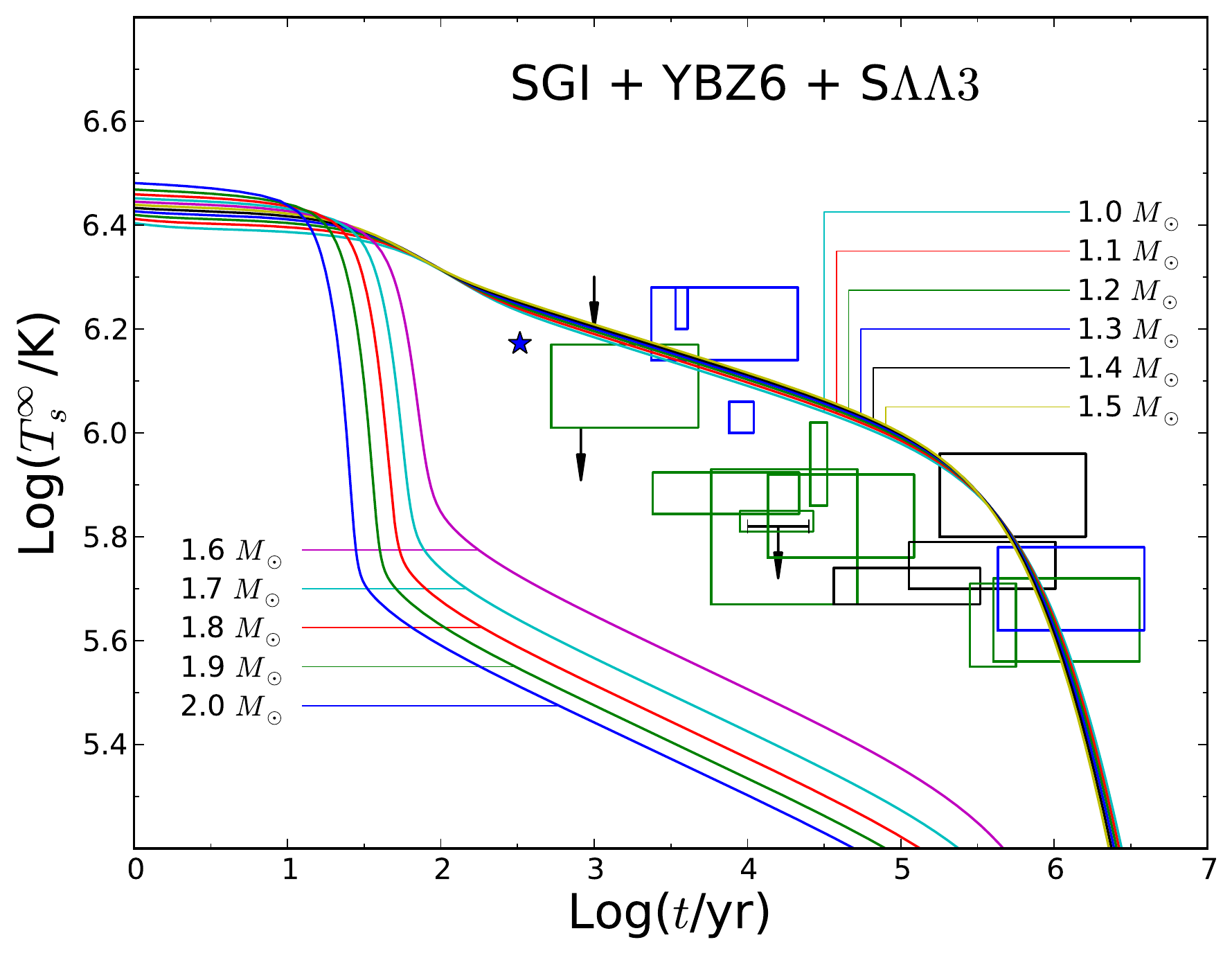}\\
(b)
\end{center}
\caption{
Cooling curves (a) without strangeness \cite{lim2015a}
and (b) with hyperons for the combined model SGI+YBZ$6$+S$\Lambda\Lambda 3$. 
}
\label{fig:sgihyper}
\end{figure}

\begin{figure}[t]
\begin{center}
\includegraphics[scale=0.4]{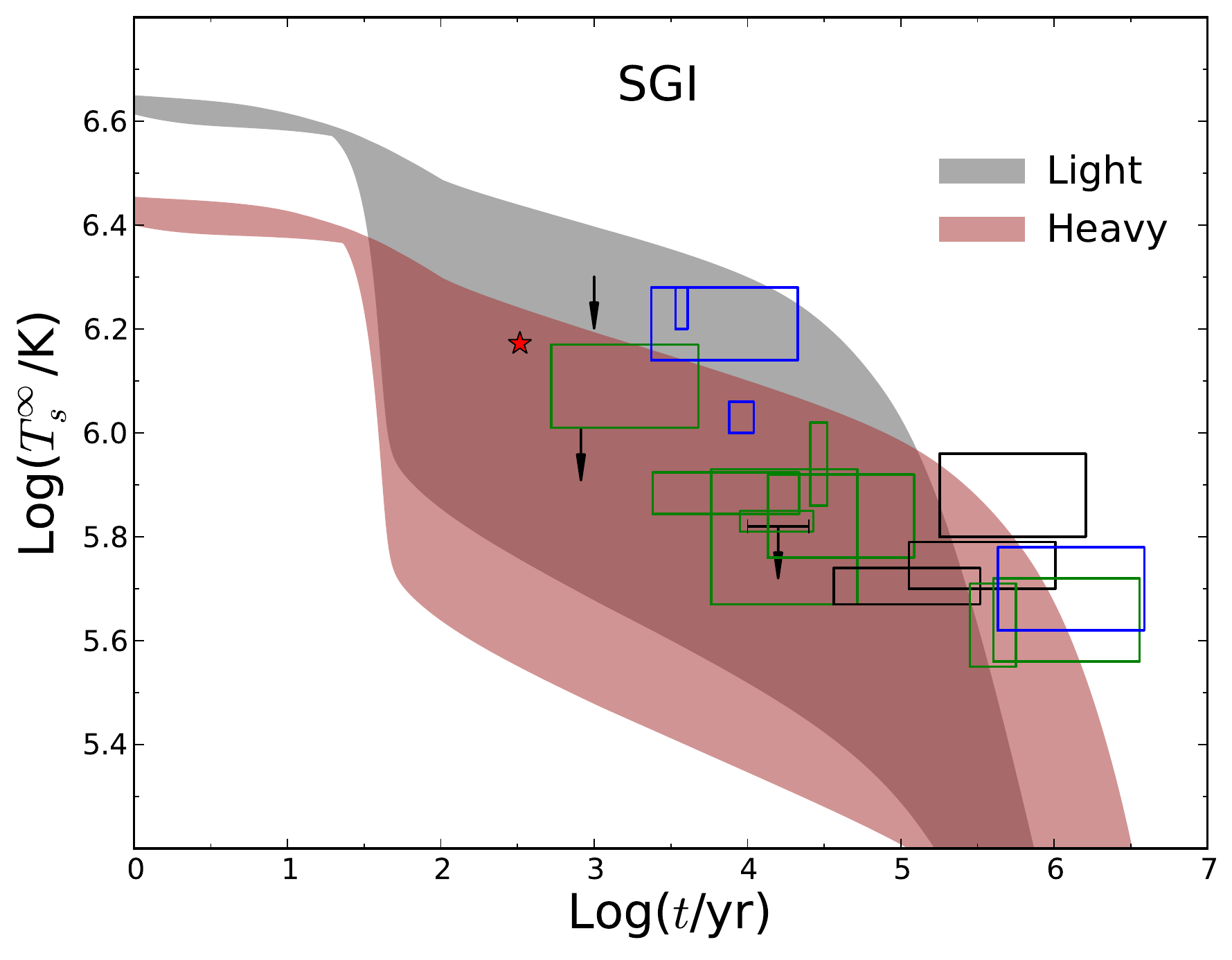}\\
(a) \\[10pt]
\includegraphics[scale=0.4]{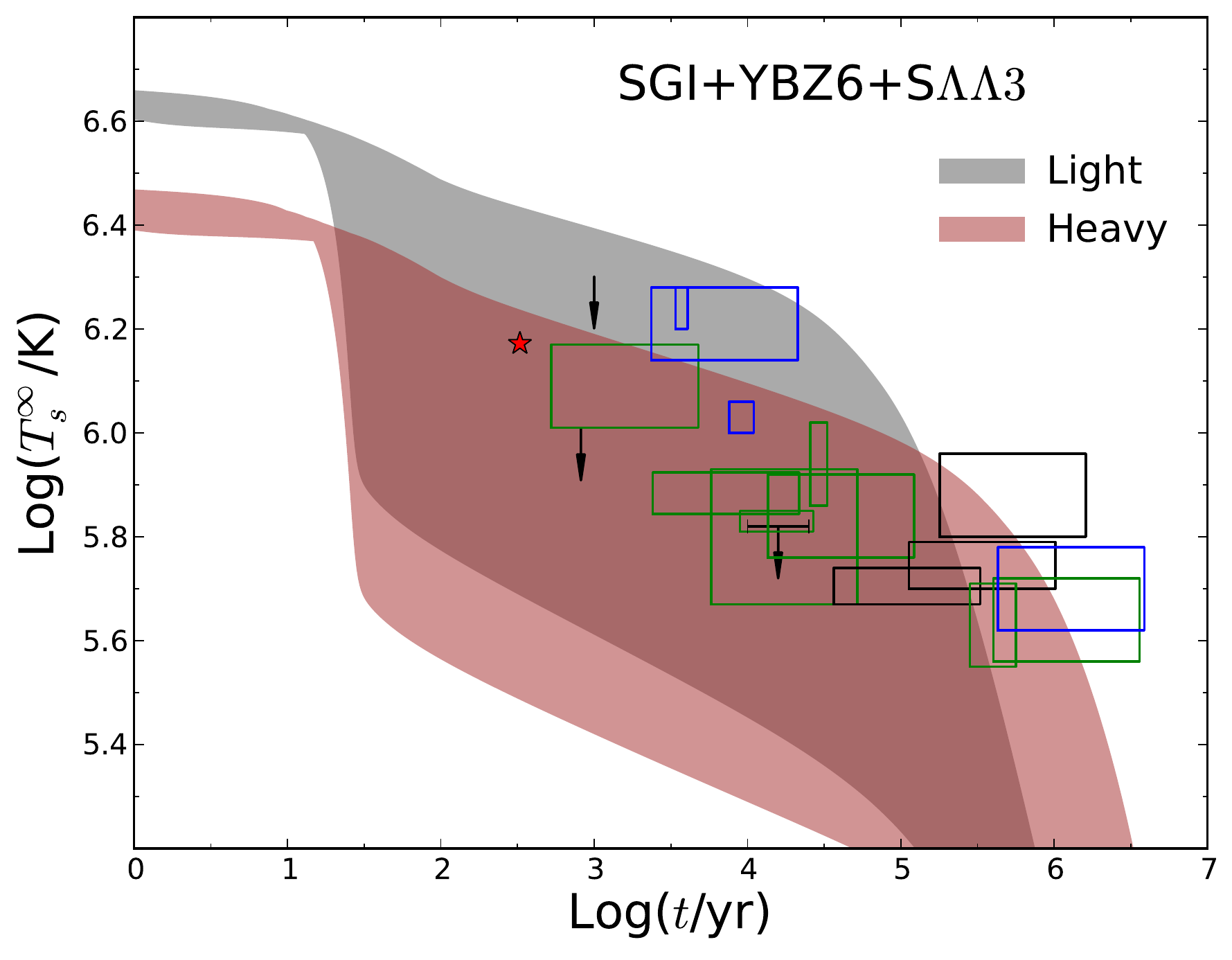}\\
(b)
\end{center}
\caption{
Dependence on the surface elements of neutron stars (a) without strangeness for SGI model \cite{lim2015a}
and (b) with hyperons,
SGI+YBZ$6$+S$\Lambda\Lambda 3$. 
In this plot, we used PCY parameterization \cite{PCY} in order to take into account both light and heavy surface elements. 
}
\label{fig:sgihypers}
\end{figure}

In Figures~\ref{fig:ski4hyper} and \ref{fig:sgihyper}, cooling curves without  and with hyperons for 
SkI4 and SGI models are shown, respectively. In these figures, we used GPE parameterization \cite{GPE} which is based on the heavy surface elements.
We also include the temperature profiles of 19 observed neutron stars
\cite{cas2010,shk2008,ztp1999,zps2004,zavlin2007,pzst2002,pzsbg2001,gowan2004,zp2004,zavlin2007a,pmc1996,hw1997,hkcap2007,pz2003,mzh2003,shsm2004a,weiss2004,hgchr2004} in all figures in this paper. 
SLy4 model is not included in these figures because the maximum mass of neutron star with hyperons
is $1.85\msun$, inconsistent with the current observation.
In these figures, by comparing plots without and with hyperons, one can see the hyperon effect is almost negligible for the neutron stars with masses below $1.5\msun$.
For both SkI4 and SGI models with hyperons, the standard cooling processes are the main cooling mechanisms
in the stars with masses below $1.5\msun$, while the stars of masses larger than $1.6\msun$
cool so fast that the temperatures are much lower than any data.
These results are reasonable in the statistical point of view since the 
most populated neutron stars have masses around $1.3 \sim 1.6 \msun$ \cite{lattimer2012}.
In addition, if the effects of envelope elements are considered, we can have broader range of
cooling curves (see \cite{lim2015a}), which may be able to include most of the observational data
with masses less than $1.6 \msun$. In Figure~\ref{fig:sgihypers}
  we summarize the dependence on the surface element of neutron stars. 
By considering the fact that the surface elements evolves from light to heavy \cite{lim2015a},
one can see that the hyperon EoS model, SGI$+$YBZ6$+$S$\Lambda\Lambda 3$, can cover most of the observational data.

In the presence of hyperons, the nucleon direct Urca process can be triggered by the $\Lambda$ hyperons 
because the fraction of neutrons can be reduced such that the momentum conservation condition
for the nucleon direct Urca can be satisfied.
In addition, the $\Lambda$ hyperon direct Urca process can occur as soon as $\Lambda$ hyperons are created. 
Hence the competition between the nucleon direct Urca and the $\Lambda$ hyperon direct Urca processes is 
important to understand the cooling curves.
For SkI4 and SGI models, the critical masses for the nucleon direct Urca process in the presence of hyperons are $1.63\msun$ and $1.64\msun$, respectively, as summarized in Table~\ref{tb:nsurca}.
Note that the critical mass of nucleon direct Urca is almost the same for SkI4 models, but it is reduced by about 5\% for SGI model due to hyperons (see Table~\ref{tb:nsmass}).

In Figure~\ref{fig:ski4hyper}, for the SkI4  model, $1.6 \msun$ neutron star cools down
mainly by the hyperon direct Urca process at the ages around $10^2$ years. 
On the other hand,
$1.7 \msun$ neutron star emits neutrinos mainly via the nucleon direct Urca process.
Thus the slope of cooling curve of $1.7 \msun$ neutron star
is stiffer than the one for $1.6 \msun$ neutron star in the early stage. 
A similar trend can be seen in the SGI model, Figure~\ref{fig:sgihyper}.

\subsection{Kaon condensation}

For SkI4 and SGI models, 
the critical mass of neutron stars for condensed kaon (lower limits in Table~\ref{tb:nskaon}) 
is greater than the one for the nucleon direct Urca processes (parentheses in Table~\ref{tb:nsmass}).
In these cases, since the nucleon direct Urca is the dominant neutrino emission process,
the cooling curves both with and without kaon condensation are almost the same as Figures~\ref{fig:ski4hyper}(a) and \ref{fig:sgihyper}(a).

\begin{figure}[t]
\begin{center}
\includegraphics[scale=0.4]{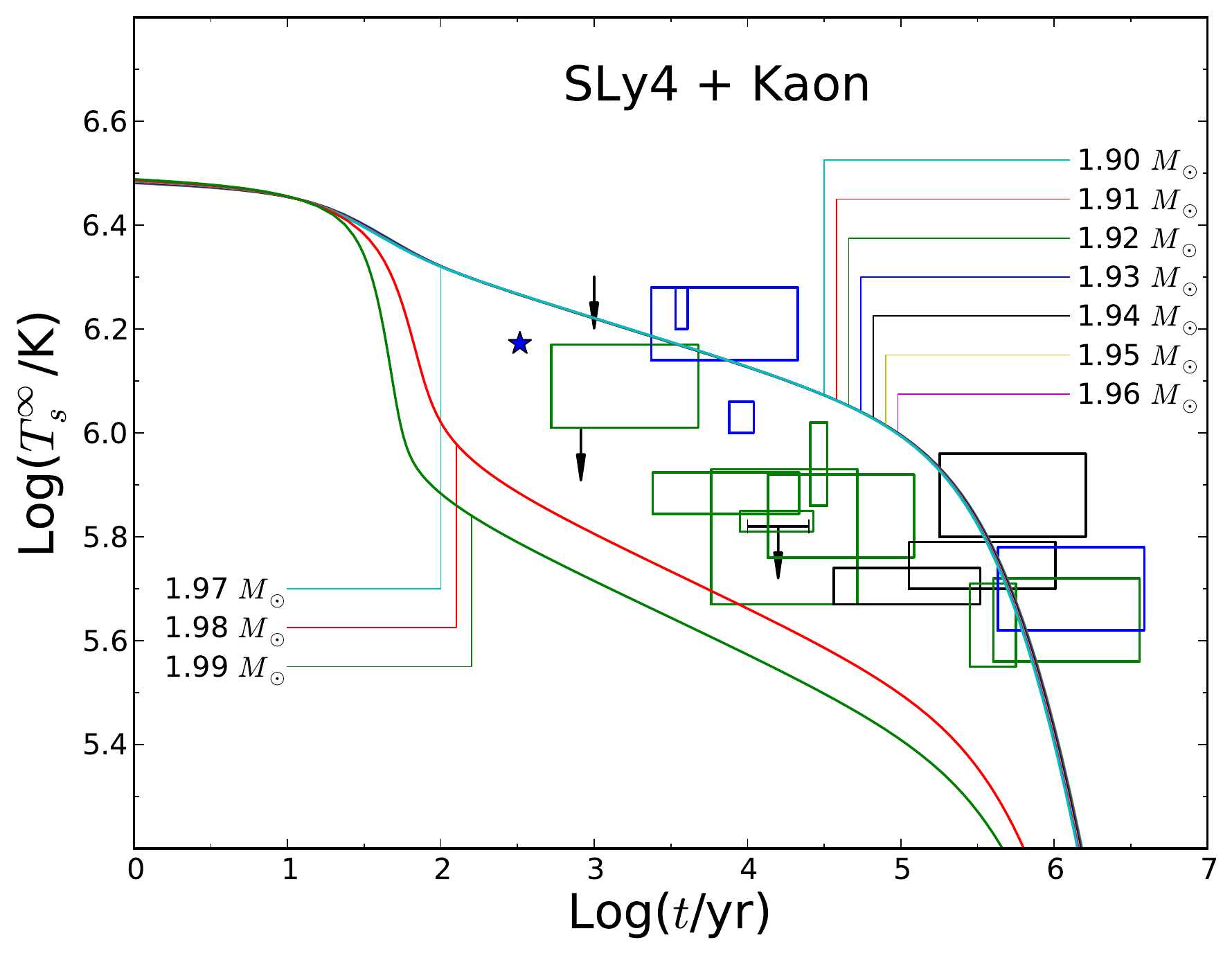}
\end{center}
\caption{Kaon condensation cooling curves from SLy4 and $a_3m_s =-134$MeV model. 
In this plot, we used GPE parameterization \cite{GPE} which is based on the heavy surface elements.}
\label{fig:sly4kaon}
\end{figure}

In Figure~\ref{fig:sly4kaon}, we plot the cooling curves of SLy4 model with $a_3m_s = -134$ MeV for 
the mass range $1.90 \sim 1.99\msun$.
For other values of $a_3 m_s$, the maximum mass is less than $2 \msun$, inconsistent with the upper
limit from the observation, so we exclude them from the consideration.
As shown in Table~\ref{tb:nsmass}~\cite{lim2015a}, with only nucleons, SLy4 model does not allow the direct Urca process
because proton fraction is not sufficient enough.
Modified Urca process is the main source of thermal evolution, so the cooling curve is insensitive to the mass
of neutron stars.
As a result, the temperature decreases very slowly,
and 
the cooling curves cannot explain mid-age low temperature
neutron stars even if we include  the effects of elements in the envelop on the surface
of neutron stars~\cite{lim2015a}.

Including kaons, the effect of kaon  (Eq.~(\ref{eq:kurca})) 
becomes apparent with the mass larger than $1.96 \msun$.
It is evident from Figure~\ref{fig:sly4kaon} that kaon condensation is a source of fast cooling.
However the effect is available only in very massive stars.
In less massive stars, cooling curve seldom differs from the one without kaon.
Therefore we can conclude that though kaon condensation can play some role to the EOS, 
it can hardly change the cooling curve of neutron stars in the most populated mass range.

\section{Discussion and Conclusion}\label{Sec:con}

The purpose of this work is to investigate the general behaviour of neutron star cooling curves 
in the presence of hyperons or kaons.

Hyperon direct Urca process may open a door to understand the thermal
evolution of neutron stars. 
We used Skyrme nuclear force models and 
non-relativistic hyperon potential models to simulate the neutron star cooling.
Both SkI4 and SGI models can satisfy the maximum mass constraint even with hyperons 
because the maximum masses without hyperons are large enough 
so that the reduced masses due to the hyperons are 
still consistent with the maximum mass constraint.
In these models, the threshold mass for the hyperon direct Urca process is smaller than the one
for the nucleon direct Urca process, and the competition between these two processes plays a crucial role 
for the cooling of neutron stars.
The presence of hyperons allows the hyperon direct Urca process with a smaller fraction of protons
compared with the nucleon direct Urca process, and the creation of $\Lambda$ hyperons can trigger the 
nucleon direct Urca process because the fraction of neutrons can be reduced.
The most drastic effects can be seen for the neutron stars with masses larger than $1.6\msun$, for which the temperature drops too early to explain the observed data.
For SGI model,  cooling curves for SGI and SGI$+$YBZ6$+$SLL3 models
are identical for the masses less than $1.44\msun$ because hyperons appear in the neutron stars with masses larger than $1.44\msun$. 
For the mass range $1.45 \sim 1.50 \msun$, the effect of hyperon is difficult to identify because the volume fraction or mass fraction with hyperon direct Urca is negligible.
For SkI4, similar conclusion can be drawn.  
Hence cooling curves of neutron stars between $1.5\msun$ and $1.6\msun$ can cover wide region in the temperature plot.
In addition, if we include the dependences on surface elements of neutron stars, the cooling curves of neutron stars with masses
in the range of $1.5\msun$ to $1.6\msun$
can cover most of the observations.

The effect of kaon condensation is most dramatic for an EOS which
does not allow the nucleon direct Urca process. 
If the superfluidity effect is not considered, 
SLy4 model without kaons gives almost identical neutron star cooling curves independently 
of neutron star masses~\cite{lim2015a}.
The presence of kaons enhances the neutrino emissivity so the cooling
curves splits if the mass of neutron stars is greater than $1.96 \msun$. 
Therefore this model shows very limited influence of kaon condensation to the cooling of neutron stars in which the standard cooling is the dominant cooling mechanism.
On the other hand, for other EOSs which allow the nucleon direct Urca process, such as SkI4 and SGI models,
the contribution from kaon neutrino emission is too small to be recognized in the cooling curve.
In these cases, the sudden drop in temperature occurs for the neutron stars with masses $1.6\msun \sim 1.7\msun$ for SkI4 models ($1.7\msun \sim 1.8 \msun$ for SGI models). Compared to the results with hyperons, the mass ranges are a bit higher to be consistent with temperature observations.

In summary, we found that the neutron star EOS with hyperons can be consistent with both mass and temperature observations. We also found that the neutron star EOS with kaons can be consistent with observations, but the cooling behavior is seldom useful to identify or isolate the effect of kaon condensation.  In order to make a firm conclusion on the effect of strangeness 
to the neutron star cooling, more work with wider range of parameters including other models is needed.

\section*{Acknowledgements}
YL was supported by the Rare Isotope Science Project of 
Institute for Basic Science funded by Ministry of Science, 
ICT and Future Planning and National Research Foundation of Korea \\
(2013M7A1A1075764). 
Work of CHH was supported by Basic Science Research Program 
through the National Research Foundation of Korea (NRF) funded 
by the Ministry of Education \\
(2014R1A1A2054096). \\
CHL was supported by the National Research Foundation of Korea (NRF) grant funded by the 
Korea government (MSIP) (2015R1A2A2A01004238 and 2016R1A5A1013277).

\end{document}